# Comparing Bibliometric Statistics Obtained from the Web of Science and Scopus

Éric Archambault

Science-Metrix, 1335A avenue du Mont-Royal E., Montréal, Québec, H2J 1Y6, Canada and Observatoire des sciences et des technologies (OST), Centre interuniversitaire de recherche sur la science et la technologie (CIRST), Université du Québec à Montréal, Montréal (Québec), Canada. E-mail: eric.archambault@science-metrix.com

David Campbell

Science-Metrix, 1335A avenue du Mont-Royal E., Montréal, Québec, H2J 1Y6, Canada. E-mail: david.campbell@science-metrix.com

Yves Gingras, Vincent Larivière

Observatoire des sciences et des technologies (OST), Centre interuniversitaire de recherche sur la science et la technologie (CIRST), Université du Québec à Montréal, Case Postale 8888, succ. Centre-Ville, Montréal (Québec), H3C 3P8, Canada. E-mail: gingras.yves@uqam.ca; lariviere.vincent@uqam.ca

**Abstract**

For more than 40 years, the Institute for Scientific Information (ISI, now part of Thomson Reuters) produced the only available bibliographic databases from which bibliometricians could compile large-scale bibliometric indicators. ISI's citation indexes, now regrouped under the Web of Science (WoS), were the major sources of bibliometric data until 2004, when Scopus was launched by the publisher Reed Elsevier. For those who perform bibliometric analyses and comparisons of countries or institutions, the existence of these two major databases raises the important question of the comparability and stability of statistics obtained from different data sources. This paper uses macro-level bibliometric indicators to compare results obtained from the WoS and Scopus. It shows that the correlations between the measures obtained with both databases for the number of papers and the number of citations received by countries, as well as for their ranks, are extremely high ($R^2 \approx .99$). There is also a very high correlation when countries' papers are broken down by field. The paper thus provides evidence that indicators of scientific production and citations at the country level are stable and largely independent of the database.

**Background and research question**

For more than 40 years, the Institute for Scientific Information (ISI, now part of Thomson Reuters), produced the only available bibliographic databases from which bibliometricians could compile data on a large scale and produce statistics based on bibliometric indicators. Though often criticized by bibliometricians (see, among several others, van Leeuwen *et al.* 2001 and Moed, 2002), Thomson's databases—the Science Citation Index (Expanded), the Social Sciences Citation Index and the Arts and Humanities Citation Index, now regrouped under the Web of Science (WoS)—were the major sources of bibliometric data until 2004, when Scopus was launched by the publisher Reed Elsevier. For those who perform bibliometric analyses and comparisons of countries or institutions, the existence of these two major databases raises the important question of the comparability and stability of statistics obtained from these two different data sources.

The comparison of these two databases has been the focus of several papers, mostly made using the "bibliographic" web versions of the databases. For instance, Burnham (2006), Bosman *et al.* (2006), Falagas *et al.* (2008), Gavel and Iselid (2008), Jacsó (2005), Neuhaus and Daniel (2008) and Norris and Oppenheim (2007) compared the general characteristics and coverage of the databases; other studies compared the bibliometric rankings obtained.[1] Given the limitations of the databases' web versions for producing bibliometric indicators, most of these studies used small samples of papers or researchers. For instance, Bar-Ilan (2008), Belew (2005), Meho and Yang (2007), Meho and Rogers (2008) and Vaughan and Shaw (2008) compared small samples of researchers' citation rates and h-indexes. Along the same line, Bakkalbasi *et al.* (2006) and Lopez-Illescas, Moya-Anegon & Moed (2008; 2009) compared citations received by a sample of journals in oncology. One of the few macro-level bibliometric studies is that of Ball and Tunger (2006), which compared the citation rates obtained with the two databases. These studies generally found good agreement between the WoS and Scopus, which is not surprising given the fact that 7,434 journals—54% of Scopus and 84% of the WoS—are indexed by both databases (Gavel and Iselid, 2008). However, they do not show whether the differences in article citation rates observed between the two databases affect the rankings of countries or institutions.

Whereas the previous papers mainly used the online version of these databases, this paper is written by licensees of these tools and is therefore based on bibliometric production platforms (implemented on Microsoft SQL Server). Using these platforms, the paper compares macro-level bibliometric indicators and provides a comparative analysis of the ranking of countries in terms of the number of papers and the number of citations received, for science as a whole as well as by fields in the natural sciences and engineering. The convergence of the bibliometric indicators will suggest that 1) the two databases are robust tools for measuring science at the country level and that 2) the dynamics of knowledge production at the country level can be measured using bibliometrics.

Using these data, the present paper, which builds on a previous abstract presented at the STI2008 conference in Vienna (Archambault, Campbell, Gingras and Larivière, 2008), examines how countries' rankings compare for both the number of papers and the number of citations. In addition to these correlation analyses based on rankings, the number of papers and the number of citations obtained in both databases at the country level are also examined. The paper then goes one step further by examining how comparable scientific output at the country level in scientific fields such as physics, chemistry and biology is. Finally, the paper examines output at the country level in the field of nanotechnology.

---

[1] More often than not, these studies also included Google Scholar. Given that this database is not yet suitable for compiling macro-level bibliometric data, this paper compares only Scopus and the Web of Science.

**Methods**

Data for this paper were produced from the WoS and Scopus databases for the 1996–2007 period. This short comparison period is a restriction imposed by Scopus, which does not include cited references prior to 1996. However, in the vast majority of cases, having the last twelve years of data is sufficient for performance measurement. Moreover, our objective is not to provide an assessment of countries but rather to compare the results obtained from the two sources in order to evaluate the robustness of the two bibliometric databases as well as of bibliometrics as a scientific undertaking.

Both bibliographic databases were received from their providers (Elsevier for Scopus and Thomson Reuters for WoS) in XML or flat files and were then transformed into relational databases implemented on SQL Server. Misspelled country names where harmonized in both databases into a preferred form, and the same form was used in both in order to match publications and citations. The categories used to delineate the fields of natural sciences and engineering are those used by the US National Science Foundation (NSF) in the production of its Science and Engineering Indicators, which is neither the original classification of the WoS nor that of Scopus[2] This taxonomy is a journal-based classification and has been in use since the 1970s. Journals that were not included in the NSF classification were manually classified. The nanotechnology datasets were built by running a query using a fairly complex set of keywords (in titles, abstracts and author keywords) in each database for the 1996-2006 period (2007 was not available at the time the data was compiled). All calculations of papers and citations use whole counting; one paper/citation is credited to each country contributing to a paper.

One of the main issues in compiling bibliometric data is the choice of the types of documents to include. In the past, bibliometricians generally used Thomson's articles, research notes and review articles, generally considered as original contributions to the advancement of science (Moed, 1995). However, since the two databases do not cover and categorize documents symmetrically, it was not possible to reproduce this selection in Scopus. Table 1 shows the differences in document types for the journal *Science* in 2000. In addition to showing that the two databases label the same documents differently, it also shows that, for document types with the same name, discrepancies are observed in document counts between the WoS and Scopus. For example, while there is a slight difference for articles, there is a significant difference for editorials, letters and reviews.

---

[2] See: http://www.nsf.gov/statistics/seind06/

Table 1. Types and number of documents in the WoS and Scopus for the journal *Science* (2000)

| Type of Document | Scopus | WoS |
| --- | --- | --- |
| Article | 906 | 859 |
| Biographical Item | | 4 |
| Book Review | | 100 |
| Conference Paper | 24 | |
| Database Review | | 1 |
| Editorial | 65 | 336 |
| Erratum/Correction | 112 | 83 |
| Letter | 217 | 322 |
| News | | 782 |
| Note | 536 | |
| Review | 170 | 61 |
| Short Survey | 446 | |
| Software Review | | 39 |

Source: Scopus data compiled by Science-Metrix, and WoS data by OST

Considering existing discrepancies in document coverage and classification between both databases, it was not possible to produce comparable subsets of documents that would match the classical set of three document types (i.e., articles, research notes and review articles). Therefore, all document types were retained when calculating the number of papers and citations in both databases, the majority of which are journal articles. This paper compares 14,934,505 papers and 100,459,011 citations received in WoS with 16,282,642 papers and 125,742,033 citations received in Scopus.

**Results**

Figure 1 compares the number of papers per country in Scopus and WoS (1a) and the countries' rankings based on these outputs (1b). The correlations between the measured outputs in both databases are remarkably strong ($R^2$ of 0.995 and 0.991 respectively). When examining top-ranking countries, Scopus often gives higher ratings to Asian countries (e.g., Japan and China each gain two ranks), whereas several European and English-speaking countries cede one rank (e.g. the U.K., Germany, France, Canada and the Netherlands). However, except for minor variations such as these, the top countries have similar ranks in both databases, the changes never exceeding a difference of two places, and the top 25 countries are the same for both databases. Figure 2 confirms that variations between the databases are quite minimal. Overall, 50% of the countries keep the same rank in both databases, 85% of the countries do not change rank by more than 5%, and 95% of the countries do not change rank by more than 10%.

Figure 1. Correlation in number of papers by country (absolute numbers and ranks), WoS and Scopus, 1996–2007

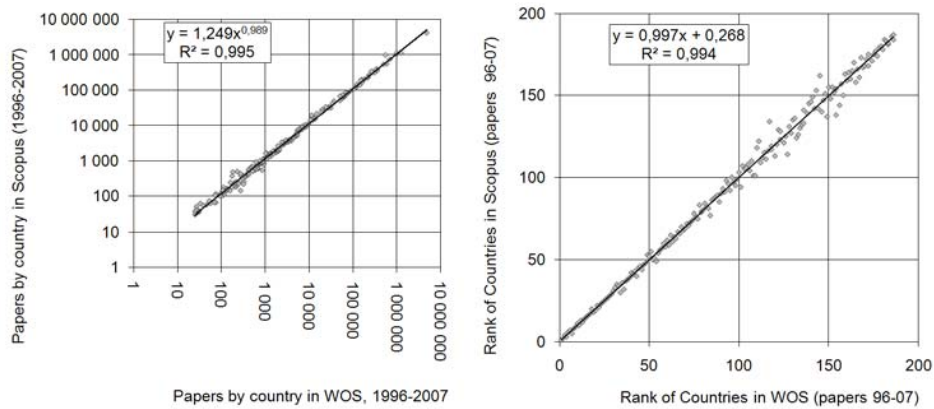

Source: Scopus data compiled by Science-Metrix, and WoS data by OST

Figure 2. Percentage of variation in countries' ranks when using WoS and Scopus, 1996–2007

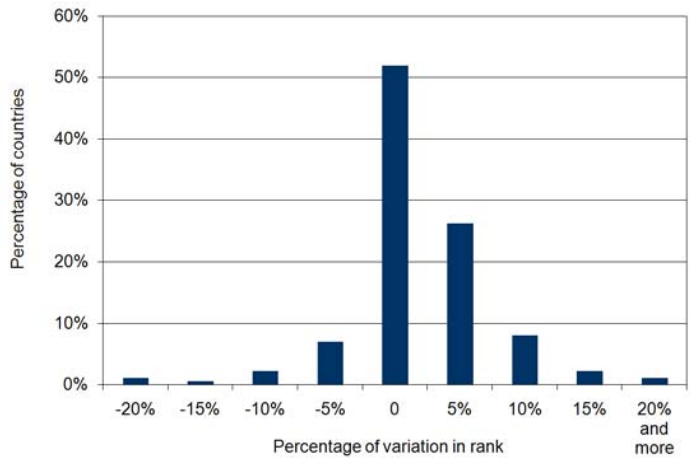

Source: Scopus data compiled by Science-Metrix, and WoS data by OST

However, these correlations might be high only because the time period considered is fairly long and the number of papers per country is therefore commensurably large. To examine the stability of the ranking with smaller datasets, the number of papers in WoS and Scopus was compared for three-year periods (Figure 3). Again, the correlation is extremely high and the $R^2$ values are consistently above the 0.99 mark. Data on ranks (not shown) are also highly correlated. This shows that country-level data for scientific output are highly similar between these two sources for science as a whole.

Although papers are an important indicator of scientific output, these data fall short of providing interesting insight into the scientific impact of nations. In this respect, citations are widely used. It is therefore relevant to ask whether citation data between these two databases are markedly different at the country level. The data presented in Figure 4 unambiguously show that countries' citation counts are extremely similar in both databases. The correlations between the two databases in terms of the countries' number of citations and ranks both have $R^2$ values above .99. The top 25 countries according to received citations are the same for both databases though there are slight variations (never exceeding two ranks) in ranking.

Figure 3. Correlation in number of papers by countries, WoS and Scopus, for three-year periods, 1996–2007

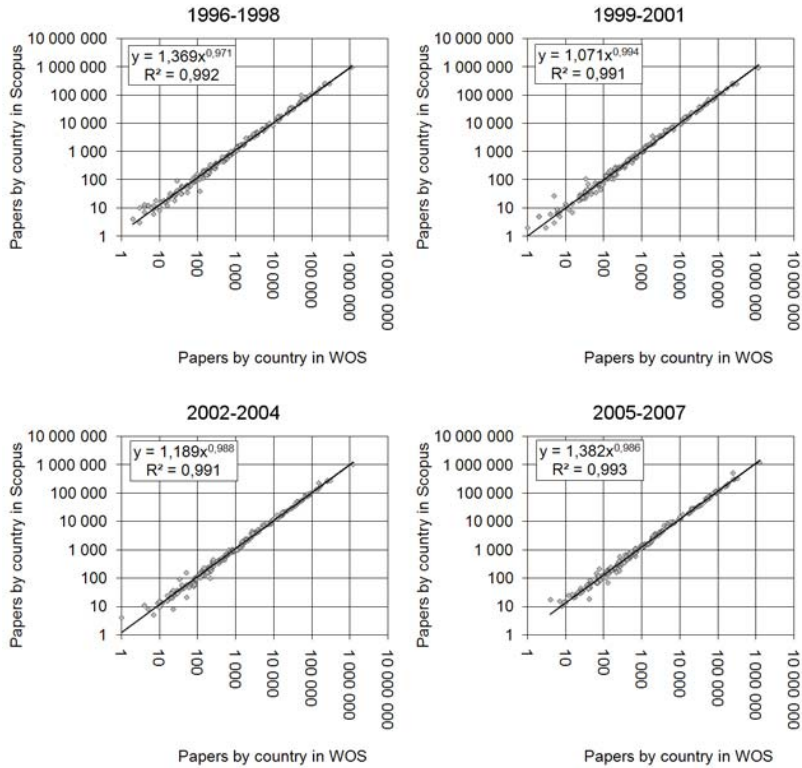

Source: Scopus data compiled by Science-Metrix, and WoS data by OST.

Figure 4. Correlation in number of citations by countries, WoS and Scopus, 1996–2007

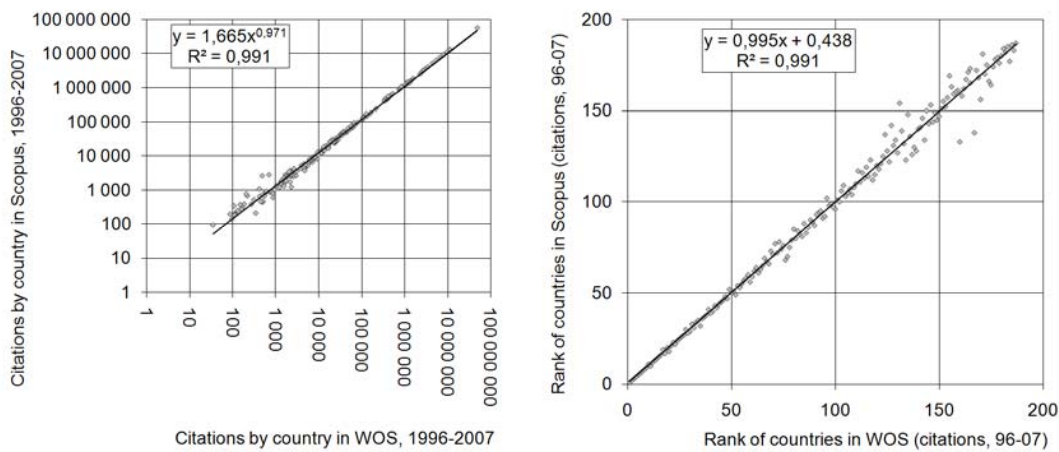

Source: Scopus data compiled by Science-Metrix, and WoS data by OST.

Finally, we computed how differently these databases measure countries' outputs in fields of the natural sciences and engineering (Figure 5) and nanotechnology (Figure 6) to examine the stability of the rankings in smaller datasets. Figure 5 shows that, in all fields except clinical medicine (.987), the correlation between the number of papers by country indexed in both databases is above .99. Even in fields where fewer papers are published (mathematics, earth and space and biology), the $R^2$ is well above .99.

The nanotechnology dataset (Figure 6) produces very similar results, the coefficient of determination ($R^2$) for the number of papers and citations being 0.991 and 0.967 respectively. Using rankings instead of absolute numbers of papers and citations, the correlations become respectively 0.990 and 0.974 (not shown). For both databases, the top 25 countries are the same in nanotechnology for both papers and citations. A few countries have somewhat different outputs in the two databases, but the databases produce remarkably similar rankings in terms of number of papers and citations for countries that have at least 100 papers. The variations for these countries never exceed six ranks for papers and seven ranks for citations.

Overall, most of the countries for which important differences were noted between the databases had either faced political turmoil that led to a breakdown (e.g., the former Yugoslavia and the U.S.S.R.) or only obtained partial recognition of their independence (e.g. a number of colonies). In the former case, divergence in the way countries were coded during transition periods in the two databases created the observed discrepancies, whereas in the latter case, papers from colonies might have been attributed differently to the colony and its mother country in the two databases (e.g. French Guyana is an "overseas department" which is considered to be an integral part of France by the French Government).

Figure 5. Correlation in number of papers by countries, WoS and Scopus, in natural sciences and engineering fields (8), 1996–2007

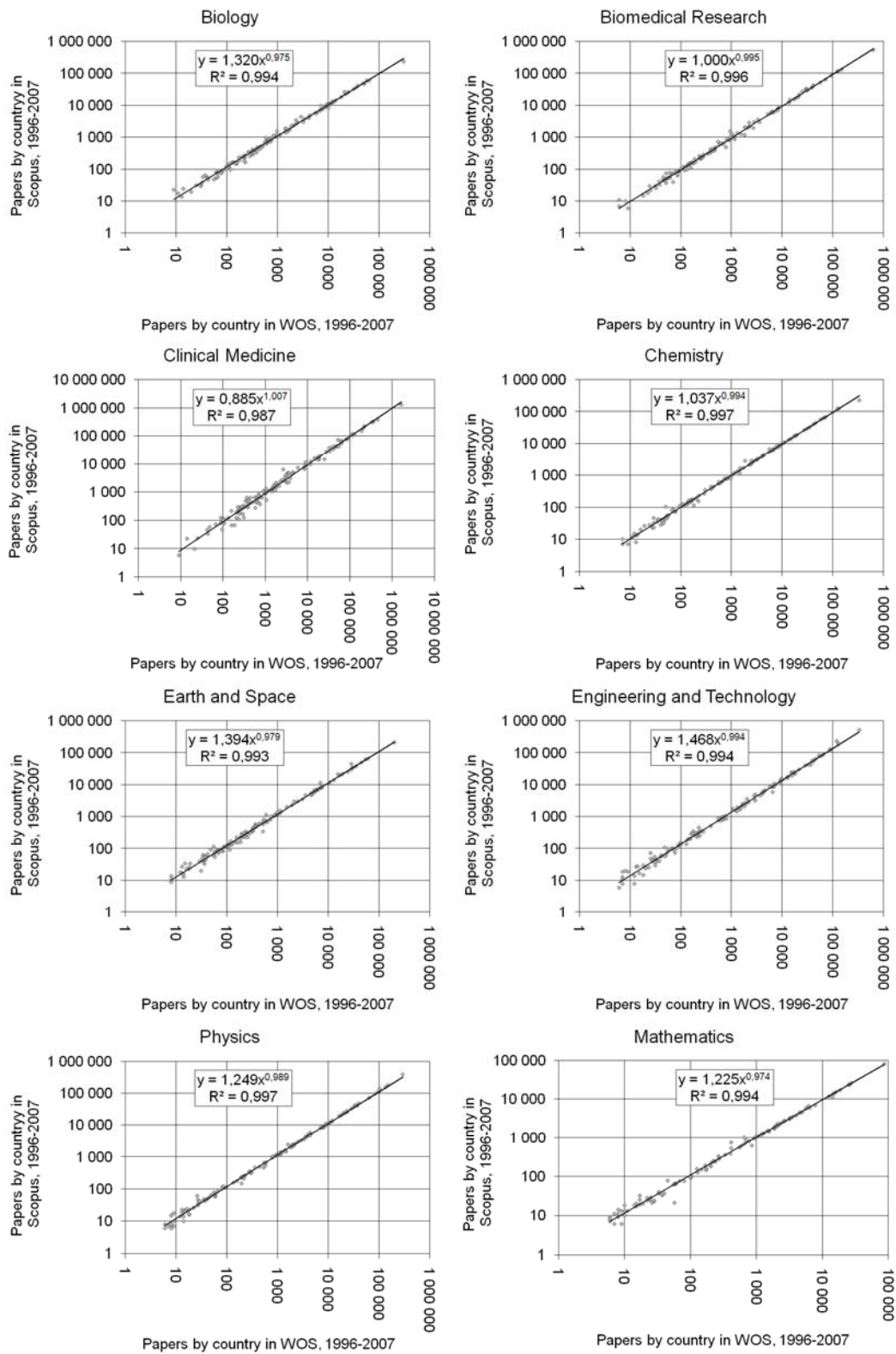

Source: Scopus data compiled by Science-Metrix, and WoS data by OST.

Figure 6. Correlation in number of papers by countries, WoS and Scopus, in nanotechnology, 1996–2006

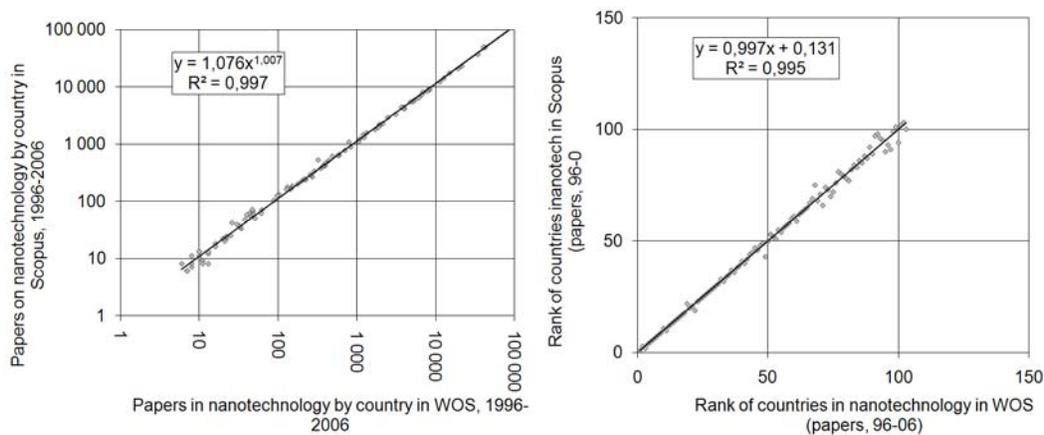

Source: Scopus data compiled by Science-Metrix, and WoS data by OST.

**Conclusion**

The above results provide strong evidence that scientometrics based on bibliometric methods is a sound undertaking at the country level. Despite the fact that the WoS and Scopus databases differ in terms of scope, volume of data and coverage policies (Lopez-Illescas, Moya-Anegon & Moed, 2008), the outputs (papers) and impacts (citations) of countries obtained from the two databases are extremely correlated, even at the level of specialties as the subsets of data in nanotechnology suggests. These results are consistent with those obtained by Lopez-Illescas, Moya-Anegon & Moed (2009) for the field of oncology. Hence, the two databases offer robust tools for measuring science at the country level. Further research using comprehensive datasets should examine differences at the institutional level as well as in different fields—such as those of the social sciences and humanities—to test whether these results still hold at lower scales.